\documentclass[manuscript]{aastex}
\slugcomment{Submitted to ApJL, March 29 2010}

\shorttitle{Binary stars in faint stellar systems}
\shortauthors{McConnachie \& C{\^o}t{\'e}}

\begin{document}

\title{Revisiting the influence of unidentified binaries on velocity
dispersion measurements in ultra-faint stellar systems}

\author{Alan W. McConnachie\email{alan.mcconnachie@nrc-cnrc.gc.ca}}
\author{Patrick C{\^o}t{\'e}\email{patrick.cote@nrc-cnrc.gc.ca}}
\affil{NRC Herzberg Institute of Astrophysics, 5071 West Saanich Road, Victoria, B.C., V9E 2E7, Canada}

\begin{abstract}

  Velocity dispersion measurements of recently discovered Milky Way
  satellites with $M_V\gtrsim-7$ imply they posses high mass-to-light
  ratios. The expected velocity dispersions due to their baryonic mass
  are $\sim0.2$\,km\,s$^{-1}$, but values $\gtrsim3$\,km\,s$^{-1}$ are
  measured. We perform Monte Carlo simulations of mock radial velocity
  measurements of these systems assuming they have mass-to-light
  ratios similar to globular clusters and posses an unidentified
  binary star population, to determine if these stars could boost the
  velocity dispersion to the observed values. We find that this
  hypothesis is unlikely to produce dispersions much in
  excess of $\sim 4.5$\,km\,s$^{-1}$, in agreement with previous
  work. However, for the systems with potentially the smallest
  velocity dispersions, values consistent with observations are
  produced in $5-40$\% of our simulations for binary fractions in
  excess of $f_{bin}(P\le10$\,yrs$)\sim5\%$. This sample includes the
  dwarf galaxy candidates that lie closest to classical globular
  clusters in $M_V-r_h$ space. Considered as a population, it is
  unlikely that all of these dwarf galaxy candidates have
  mass-to-light ratios typical of globular clusters, but boosting of
  the observed dispersion by binaries from near-zero values cannot be
  ruled out at high confidence for several individual dwarf galaxy
  candidates. Given the importance of obtaining accurate velocity
  dispersions and dynamical masses for the faintest satellites, it is
  clearly desirable to exclude directly the possible effect of
  binaries on these systems. This requires multi-epoch radial velocity
  measurements with individual uncertainties of
  $\lesssim$1\,km\,s$^{-1}$ to identify spectroscopic binaries with
  orbital velocities of order the observed velocity dispersion.
\end{abstract} 


\section{Introduction}

The internal velocity dispersions of Milky Way (MW) dwarf galaxies
with $M_V\le-8$ lie in the range $6<\sigma_{obs}<11$\,km\,s$^{-1}$
\citep{aaronson1983}. These values are large given the observed
luminosities and surface brightnesses, and prompted early research to
consider unidentified spectroscopic binary stars as a potential
explanation for the unexpectedly high velocity dispersions (e.g.,
\citealt{mateo1993,olszewski1996,hargreaves1996}). If many of the
targeted stars are binaries, then the radial velocity that is measured
is a superposition of their intrinsic radial velocity within the
galaxy and their binary orbital velocity. Simulations of plausible
binary populations showed that, while some inflation of the
dispersion may be expected due to binaries, this effect is
generally insufficient to explain the high velocity dispersions that
are measured for the dwarf galaxies. More recently, \cite{minor2010}
develop a methodology to determine the contribution to the velocity
dispersion of a dwarf galaxy from a binary population with a given set
of parameters, for a well sampled dataset.

Here we revisit the role of binary stars in boosting the observed
velocity dispersion of intrinsically low-mass systems. We are
motivated by the discoveries of MW satellites with considerably lower
luminosity ($M_V\gtrsim-7$) than previously known dwarf galaxies, a
regime that was not considered in the original papers on this
subject. Some of these systems have few bright stars suitable for
spectroscopic follow-up, and a few have measured velocity dispersions
as low as $\sigma_{obs}\sim3$\,km\,s$^{-1}$ (implying mass-to-light
ratios $\Upsilon_V \equiv M/L_V \gtrsim 100$). In Section~2, we
summarize the relevant properties of these satellites, and we discuss
how we generate realistic binary populations. In
Section~3, we use Monte-Carlo simulations to investigate the effect of
binary stars on velocity dispersion measurements of these systems
assuming they have a near-zero intrinsic dispersion.

\section{Preliminaries}

\subsection{The faint population of Milky Way satellites}

Fourteen new MW satellites discovered since 2004
(\citealt{willman2005,willman2006,zucker2006a,zucker2006b,belokurov2006b,belokurov2007,belokurov2008,belokurov2009,sakamoto2006,irwin2007,walsh2007})
have been shown to possess velocity dispersions significantly higher
than expected if they have mass-to-light ratios typical of globular
clusters in virial equilibrium\footnote{\cite{belokurov2010} recently
  discovered Pisces~II, a candidate MW dwarf galaxy that does not yet
  have a velocity dispersion measurement.}
(\citealt{martin2007,simon2007,geha2009,koch2009,walker2009,aden2009,belokurov2009}). Their
luminosities span more than five magnitudes, with the brightest
systems (Canes Venatici~I and Leo~T) of comparable luminosity to Draco
(the faintest MW dwarf galaxy known previously).

Table~1 summarizes the relevant structural and kinematic data for
recently discovered faint MW satellites. We define this as
$M_V\gtrsim-7$, and note that this sample includes three systems (Ursa
Major~I,~II and Bootes~I) with velocity dispersions of the same order
as those of previously known dwarfs. In addition to the observed
velocity dispersion ($\sigma_{obs}$), we list the number of putative
member stars each measurement is based on ($n_{stars}$), and the
median velocity uncertainty ({\~v}$_{err}$) of those members. Leo~V
has two dispersion measurements, a lower central value using five
stars and a larger global value using an additional two stars located
at larger radius. Here, we quote the larger value. Further,
\cite{walker2009} do not rule out that Leo~V may be a star cluster
since its dispersion is presently barely resolved. We do not include
Bootes~III ($\sigma_{obs}\simeq14$\,km\,s$^{-1}$ ) in Table~1;
\cite{grillmair2009} and \cite{carlin2009} both suggest that
Bootes~III is a dwarf galaxy undergoing tidal disruption. While we
confine our analysis to the MW satellite population, we note that a
similar analysis can also be applied to other Local Group satellites,
such as the M31 satellite system, for which kinematic data is now
becoming available (e.g., \citealt{kalirai2010,collins2010}).

For each satellite in Table~1, we calculate what their properties
would be if they possess structural and kinematic characteristics
typical for globular clusters (whose mass-to-light ratios are
consistent with expectations from population synthesis models, without
dark matter). Specifically, we derive their concentration, $c$, and
core radius, $r_c$ using the relations of \cite{mclaughlin2000}, and
derive the corresponding intrinsic velocity dispersion,
$\sigma_{int}$, using

\begin{equation}
\sigma_{int}=\left(\frac{2\pi G\alpha p M}{9\nu r_c}\right)^{\frac{1}{2}} 
\end{equation}

\noindent assuming $M=\Upsilon_VL_V$ where $\Upsilon_V=1.45\pm0.1$ in
solar units \citep{mclaughlin2000}. The constants $\nu$, $\alpha$ and
$p$ are calculated by interpolation of the values given in
\cite{king1966} and \cite{peterson1975}. Results are shown in
Table~2. The uncertainty on $\sigma_{int}$ corresponds to the
uncertainty on $\Upsilon_V$. The derived values for
$\sigma_{int}$ are $\sim0.2$\,km\,s$^{-1}$, generally at least an
order of magnitude smaller than $\sigma_{obs}$ and {\~v$_{err}$.

\subsection{Calculating the velocity dispersion}

We will derive velocity dispersion estimates from mock radial velocity
catalogs for these satellites by attempting to reproduce the
observational techniques used to measure $\sigma_{obs}$. In general,
this involves identifying stars belonging to the system and
measuring their mean velocity and dispersion. Here, we
know {\it a priori} that our stars are members, but we will not
necessarily use all the stars in our mock catalog to measure the
dispersion. For example, some short-period binaries may have very
different velocities from the mean of the sample, and they
may not be observationally identified as members of the
satellite. Instead, we sigma-clip our samples at 3$\sigma$ in order to
mimic observations as closely as possible (e.g.,
\citealt{simon2007,martin2007}). At most, this removes only 1 or 2
stars from the final sample. All of the putative member stars
identified in the actual observations summarized in Table~1 lie within
3$\sigma$ of the mean velocity, and so this seems reasonable.

Several of the radial velocity studies listed in Table~1 calculate the
velocity dispersion ($\sigma_P$) and mean velocity ($\bar{u}$) of the
putative members by maximizing the natural logarithm of the
probability function,

\begin{eqnarray}
\ln(p)& = &-\frac{1}{2}\Sigma_{i=1}^{n_{star}}\ln\left(v_{i,err}^2+\sigma_p^2\right)-\frac{1}{2}\Sigma_{i=1}^{n_{star}}\frac{(v_i-\bar{u})^2}{\left(v_{i,err}^2+\sigma_p^2\right)}\nonumber\\
&&-\frac{n_{stars}}{2}\ln(2 \pi)
\end{eqnarray}

\noindent (\citealt{walker2006a}), where $v_{i,err}$ is the
measurement uncertainty on the measured velocity, $v_i$, of the
$i^{th}$ star. We use this technique to calculate the velocity
dispersion of our mock datasets for every satellite. We test this
method on the observed datasets, and derive the same central values as
the published results, with the exception of Segue~II and Leo~V. For
Segue~II, \cite{belokurov2009} use a different likelihood analysis;
for Leo~V, \cite{walker2009} use a similar technique to Equation~2 but
incorporate the probability that a star is a member of the system. To
ensure consistency, we recalculate the velocity dispersions for
Segue~II and Leo~V using Equation~2 and use these new values in our
subsequent analysis. These are included in Table~1 in parentheses,
with the uncertainties calculated following \cite{walker2006a}.

\subsection{The binary star population}

The observed radial velocity of the primary companion in a binary star
system is given by

\begin{equation}
V_r=\frac{2\pi a_1\sin\,i}{P (1-e^2)^{\frac{1}{2}}}\left[\cos(\theta+\omega)+e\cos\omega\right]~.
\end{equation}

\noindent $a_1$ corresponds to the semi-major axis of the
relative orbit of the primary; $P$ is the period of the orbit; $e$ is the eccentricity;
$i$ is the orbital inclination to our line-of-sight; $\theta$ is
the phase of the binary measured from periastron; $\omega$ is the
longitude of periastron; $m_1$ and $m_2$
are the masses of the primary and secondary, respectively. $m=m_1+m_2$
and the mass ratio $q=m_2 /m_1$. For an equal-mass
binary system on a circular orbit with $m=1\,M_\odot$ and $P=1\,$yr,
the maximum radial velocity that can be observed is
$V_{r,max}\simeq\pm15$\,km\,s$^{-1}$. For $P=[10,100,1000]$\,yrs,
$V_{r,max}\simeq[\pm7,3,1.5]$\,km\,s$^{-1}$. 

We now summarize the adopted distributions for each of the variables
in Equation~3.  At the present time, the relevant physical parameters
are entirely unknown for the stellar systems under discussion, and so
we must be guided in our choice by the findings from other stellar
populations, particularly the solar neighborhood (SN) and halo globular
clusters.

\subsubsection{Mass distributions}

We set the mass of the primary to be $m_1=0.8\,M_\odot$; this is the
approximate minimum mass of a star that has had time to evolve on to
the giant branch. The studies summarised in Table~1 target either
giants or stars near the main-sequence turn-off, and so this seems
reasonable.

\cite{duquennoy1991} show that the mass ratio distribution for binary
stars in the SN is well approximated as log-normal,

\begin{equation}
\frac{dN}{dq} \propto \exp\left[-\frac{\left(q - \bar{q}\right)^2}{2\sigma^2_{q}}\right]~.
\end{equation}

\noindent $\bar{q} = 0.23$ and $\sigma_{q} = 0.42$. We set $q_{min} =
0.1$, so that the minimum mass of the secondary approximately corresponds to
the limiting mass for hydrogen burning.

\subsubsection{Period distribution}

We use two different binary period distributions. We first follow
\cite{duquennoy1991}, who show that the distribution for SN binaries
is described by a log-normal,

\begin{equation}
\frac{dN}{dlog_{10}P}\propto\exp\left[-\frac{\left(log_{10}P-\overline{log_{10}P}\right)^2}{2\sigma^2_{log_{10}P}}\right]~.
\end{equation}

\noindent $\overline{{log_{10}P}}=4.8$ and $\sigma_{log_{10}P}=2.3$,
where $P$ is measured in days. We also explore a uniform distribution
in $log_{10}P$.

The maximum possible period of a binary star in a stellar cluster,
$P_{max}$, is set physically by the maximum separation beyond which
the binary becomes unbound due to stellar interactions within the
cluster. \cite{hills1984} derive the semi-major axis at which
this occurs,

\begin{equation}
a_{max}=12.4AU\left(\frac{m_1+m_2}{1.4M_\odot}\right)\left(\frac{10~km\,s^{-1}}{\sigma}\right)^2
\end{equation}

\noindent where $\sigma$ corresponds to the intrinsic velocity
dispersion of the cluster. For the values of $\sigma_{int}$ in Table~2,
this corresponds to $P_{max}\sim10^7$\,yrs, or $V_{r,max} \sim
0.14$\,km\,s$^{-1}$, of order $\sigma_{int}$. However, such systems,
if present, are likely very rare (following Equation~5) and cannot
contribute to inflating the velocity dispersion by more than a small
fraction of a km\,s$^{-1}$.

Constraints on $P_{min}$ can be derived physically for red giant
branch (RGB) stars from mass transfer considerations,

\begin{equation}
a_{min}=\frac{R_1}{h(q)}~,~~0.38\lesssim h(q)\lesssim0.59~,
\end{equation}

\noindent where $R_1$ is the radius of a star on the RGB and $h(q)$ is
a slowly varying function that gives the volume radius of the Roche
lobe in terms of the orbital separation (see \S{V} of
\citealt{pryor1988}). In old, metal-poor populations, luminous RGB
stars have $R_1\sim0.5$~AU, and so we conservatively adopt
$\langle{a_{min}}\rangle \sim 1$~AU. Thus, $P_{min}\sim1$\,yr
typically, although values as low as $0.1$\,yr are plausible,
particularly for stars fainter than the tip of the RGB.

\subsubsection{Eccentricity distribution}

Two eccentricity distributions are used. Following \cite{heggie1975}, we
first assume a thermal distribution in eccentricity,

\begin{equation}
\frac{dN}{de}\propto2e~.
\end{equation}

\noindent In the second instance, we assume circular orbits throughout
($e=0$)

\subsubsection{Angles}

The probability of observing a binary system at an inclination
$i$ is proportional to $sin\,i$. The probability of observing a binary
system with a given orientation of its major axis to the line of
sight, $\omega$, is uniform between $0$ and $\pi$. The probability of
observing a binary system at a given phase, $\theta$, is proportional
to the inverse of the angular velocity, $\dot{\theta}^{-1}(m, P,
e)$. Thus, for circular orbits, the probability of observing a binary
star at a given $\theta$ is uniformly distributed. For elliptical
orbits, the binary star is more likely to be observed near
apastron. 

\subsubsection{Binary fractions}

The fraction of stars that are binaries in the systems under
discussion ($f_{bin}$) is presently entirely unconstrained, and there
is some uncertainty as to the binary fraction in other stellar
populations. For example, in the solar neighborhood,
\cite{duquennoy1991} find an overall binary fraction of $\sim$2/3 for
stars near the middle of the main sequence and \cite{hut1992} conclude
that the binary fraction in globular clusters is only slightly smaller
than that in the Galactic disk. Few constraints exist in dwarf
galaxies; \cite{olszewski1996} estimate a binary fraction in Draco and
Ursa Minor of $0.2 - 0.3$ per decade in period, for periods around
1\,yr, that is potentially $3 - 5$ times higher than in the solar
neighborhood, but which is also highly uncertain. However,
\cite{lada2006} argue that most stars are in fact single, with
two-thirds of main sequence systems in the Galactic disk composed of
single stars. In a small sample of globular clusters, \cite{cote1996}
have estimated a binary fraction of $\sim0.01-0.08$ per decade of
period (with significant error bars), and other studies such as
\cite{davis2008} in NGC~6397 imply a present-day binary fraction of
order a percent. Formation models of globular clusters can create
primordial binary fractions of 5\% (\citealt{hurley2007}) to 100\%
(\citealt{ivanova2005}). In short, it is presently unclear what the
primordial or current binary fraction of the systems listed in Table~1
will be.

Given the lack of constraints on $f_{bin}$, we examine three different
possible normalizations, $f_{bin}(P\le10$\,yrs$)=0.05, ~0.15$ and
$0.30$. The latter normalization is high, but is similar in magnitude
to that estimated by \cite{olszewski1996}. The period distribution must
be truncated at some upper limit else overall binary fractions $> 1$
can be obtained depending on the shape of the assumed distribution;
since we are only interested in binaries numerous enough and with
short-enough periods to contribute to $\sigma_{obs}$, we do not
consider periods in excess of $10 - 100$\,yrs.

\section{Analysis}

\subsection{Method}

We simulate four different binary populations with different
assumptions about the period and eccentricity distributions: (i)
log-normal in $P$, thermal in $e$; (ii) log-normal in $P$, $e=0$;
(iii) uniform in $log_{10}P$, thermal in $e$; (iv) uniform in
$log_{10}P$, $e=0$. In each case, we generate $10^6$ binary star
systems and output the results to a file.

For each of the binary populations considered, we
generate mock velocity datasets for each satellite using
the following procedure:

\begin{enumerate}

\item For $i=1,...n_{stars}$, we select an intrinsic velocity $v_i$
  for each star from a Gaussian distribution with width $\sigma_{int}$;

\item For each star, we randomly designate it as single or binary according to $f_{bin}$;

\item For stars that are designated binary, we randomly assign them
  one of the $10^6$ binary orbital properties generated
  previously (ensuring $P_{min}\le P\le P_{max}$). A new value for
  $v_i$ is obtained by summing the previous value and the orbital
  radial velocity of the primary;

\item For each star, we assign it an observational uncertainty from
  the list of $n_{star}$ observational uncertainties ($v_{err,i}$)
  that were actually obtained through observations for that satellite.
  These are available in the corresponding papers referenced in
  Table~2 or were provided by the authors (J. Simon \& M. Geha, {\it
    private communication}). $v_i$ is then modified by an error-term
  selected from a Gaussian distribution of width $v_{err,i}$;

\item Once $n_{star}$ mock radial velocities are generated according
  to $(1)-(4)$, we make a first estimate of the mean
  ($\frac{\Sigma_i^{n_{star}} v_i}{n_{star}}$) and standard deviation
  ($\sqrt\frac{\Sigma_i^{n_{star}} (v_i-{\bar v})^2}{n_{star}-1}$). All
  stars that are more than 3 standard deviations from the mean
  velocity are not considered further;

\item For the sigma-clipped sample of stars, we calculate and record
  the final values for the mean velocity and velocity dispersion using
  Equation~2;

\item Steps $(1)-(6)$ are repeated 10\,000 times for each satellite.

\end{enumerate}

\noindent We record the percentage of times the final velocity
dispersion estimate is greater than or equal to the lower 1$\sigma$
error bound of $\sigma_{obs}$. These percentages are listed in Table~3
for each of the binary populations considered, for
different values of $P_{min}$ and $P_{max}$, assuming different
binary fractions. Figure~1 shows examples of the distribution of
measured velocity dispersions from these simulations for Ursa Major~I
(left panel), Segue (middle panel) and Leo~IV (right panel). In each
panel, we show the results corresponding to $f_{bin}(P\le
10$\,yrs$)=0.15$, a log-normal binary period distribution ($1\le P\le
100$\,yrs) and circular orbits (column~7 of Table~3).

\subsection{Results}

Table~3 shows that the binary populations considered
here are very likely insufficient to explain the velocity dispersion
measurements for Ursa Major~I, II and Bootes~I (the three systems with
largest $\sigma_{obs}$; see the left panel of Figure~1), and similarly
for Coma Berenices and Canes Venatici~II (unless there is a high
binary fraction in these systems).

For the remaining systems in Table~3 (Segue~I, II, Willman~I,
Bootes~II, Leo~IV, V and Hercules), there is a modest, but
non-negligible, chance ($\gtrsim 5\%$) that the measured values of
$\sigma_{obs}$ may be due to binary stars, unless the binary fraction
is $\lesssim5\%$. For higher binary fractions, the probability that
binary stars can produce $\sigma_{obs}$ can increase to
$\gtrsim20\%$. All of the satellites most affected have
$\sigma_{obs}\lesssim4.5$km\,s$^{-1}$ (with the exception of
Bootes~II, but here the large observational uncertainty allows a
value as low as 3.1km\,s$^{-1}$). The middle panel of Figure~1 shows
an example velocity dispersion distribution for Segue, a fairly
typical case. Leo~IV is the satellite potentially most affected by
binaries (right panel of Figure~1), and inspection of Table~1 shows
that this is a result of a relatively low value of $\sigma_{obs}$ in
combination with a relatively large {\~v}$_{err}$.

Given that we do not know the actual binary fractions in any of the
systems listed in Table~1, or the distribution of their properties, we
conclude that the present observational data on the satellites with
lowest $\sigma_{obs}$ are unable to exclude at high confidence that
their velocity dispersions may result from binary stars in intrinsically
low-mass systems. On the other hand, $\sigma_{obs}$ for Ursa~Major~I
is extremely unlikely to be due to binary stars, and results for
Ursa~Major II, Bootes~I, Coma Berenices and Canes~Venatici~II are
unlikely to be able to be explained by binary stars unless these
systems possess high binary fractions. Considered as a population, it
is unlikely that all these systems have mass-to-light ratios
typical of globular clusters, but boosting of the observed dispersion
by binaries from near-zero values cannot be ruled out at high
confidence for several individual dwarf galaxy candidates. 

In line with earlier studies that dealt with higher velocity
dispersion dwarfs (\citealt{mateo1993,olszewski1996,hargreaves1996}),
we find binary stars cannot boost velocity dispersions to values much
in excess of 4.5km\,s$^{-1}$. Our results are also consistent with
those of \cite{minor2010}, who examine the effect of binaries on dwarf
galaxies with $\sigma_{int}\gtrsim4$\,km\,s$^{-1}$, which is an order
of magnitude larger than the values of $\sigma_{int}$ considered here.

The binary stars that contribute to boosting the
velocity dispersion from $\sigma_{int}$ to $\sigma_{obs}$ must have
current orbital radial velocities of order $\sigma_{obs}$. In our
simulated data, we find that the average yearly change in the orbital
radial velocities of such stars is $\sim1-3$\,km\,s$^{-1}$. To
identify such stars, multi-epoch measurements (with a cadence of a
year) with individual velocity uncertainties $\lesssim1$km\,s$^{-1}$
would be required; these observations will then directly determine the
possible influence of binary stars on $\sigma_{obs}$.

\section{Summary}

We have examined the possibility that unidentified binary stars may
inflate the velocity dispersions measured for some low-mass MW
satellites. If such objects were to have mass-to-light ratios similar
to globular clusters, then they would be expected to have
intrinsic velocity dispersions of
$0.1\lesssim\sigma_{int}\lesssim 0.3$km\,s$^{-1}$ --- significantly
below the precision of existing radial velocity measurements in these
systems. Our Monte Carlo simulations reveal that, although binaries
are unlikely to explain satellite velocity dispersions much in excess
of $\sim$4.5 km/s, boosting of the observed dispersion by binaries
cannot be ruled out with high confidence for some faint satellite
candidates. Figure~2 --- which plots $r_h$ versus $M_V$ for MW globular
clusters, dwarf galaxies and dwarf galaxy candidates --- shows that it
is those dwarf candidates with $r_h<70$\,pc (i.e. lying nearest to
the globular cluster population) that are the most susceptible to
contamination by spectroscopic binaries. Given the importance of
obtaining accurate velocity dispersions and dynamical masses for the
faintest MW satellites, it is clearly desirable to exclude directly
the possible effect of binaries on these systems by obtaining
multi-epoch radial velocity measurements with individual uncertainties
of $\lesssim$1\,km\,s$^{-1}$.

\acknowledgements

We thank Josh Simon and Marla Geha for sending us their data, and
Peter Stetson and Nicolas Martin for a careful reading of the manuscript.

\newpage

{\rotate

\begin{table*}
\hspace{-3.5cm}
\begin{tabular*}{1.2\textwidth}{l|ccccc|l}
                  & $M_V$ & $r_h$ (pc) & $\sigma_{obs}$ (km\,s$^{-1}$) & {\~v}$_{err}$ (km\,s$^{-1}$) & $n_{stars}$ & References\\
\hline
&&&&&\\
Segue             & $-1.5^{+0.6}_{-0.8}$&  $29^{+8}_{-5}$   &  $4.3 \pm 1.2$       & 5.2 & 24 & \cite{martin2008,geha2009}\\
Segue II          & $-2.5 \pm 0.3$      &  $34 \pm 3$       &  $3.4^{+2.5}_{-1.2}$ & 1.2 &  5 & \cite{belokurov2009}\\
& &  &  ($2.9 \pm 1.1$) &  & & \\
Willman I         & $-2.7 \pm 0.7$      &  $25^{+5}_{-6}$   &  $4.3^{+2.3}_{-1.3}$ & 4.4 & 14 & \cite{martin2008}; \cite{martin2007}\\
Bootes II         & $-2.7 \pm 0.9$      &  $51 \pm 17$      &  $10.5 \pm 7.4$      & 4.1 &  5 & \cite{martin2008,koch2009}\\
Coma Berenices    & $-4.1 \pm 0.5$      &  $77 \pm 10$      &  $4.6 \pm 0.8$       & 4.9 & 59 & \cite{martin2008,simon2007}\\
Ursa Major II     & $-4.2 \pm 0.5$      & $140 \pm 25$      &  $6.7 \pm 1.4$       & 4.5 & 20 & \cite{martin2008,simon2007}\\
Leo V             & $-4.3 \pm 0.5$      &  $42 \pm 6$       &  $3.7^{+2.3}_{-1.4}$ & 2.1 &  7 & \cite{belokurov2008,walker2009}\\
& & &  ($3.3 \pm 1.3$) &  &  & \\
Canes Venatici II & $-4.9 \pm 0.5$      &  $74^{+14}_{-10}$ &  $4.6 \pm 1.0$       & 2.9 & 25 & \cite{martin2008,simon2007}\\
Leo IV            & $-5.0^{+0.6}_{-0.5}$& $116^{+26}_{-34}$ &  $3.3 \pm 1.7$       & 5.2 & 18 & \cite{martin2008,simon2007}\\
Ursa Major I      & $-5.5 \pm 0.3$      & $318^{+50}_{-39}$ &  $7.6 \pm 1.0$       & 3.4 & 39 & \cite{martin2008,simon2007}\\
Bootes I          & $-6.3 \pm 0.2$      & $242^{+22}_{-20}$ &  $6.5^{+2.0}_{-1.4}$ & 2.7 & 30 & \cite{martin2008}; \cite{martin2007}\\
Hercules          & $-6.6 \pm 0.3$      & $330^{+75}_{-52}$ &  $3.72 \pm 0.91$     & 3.0 & 18 & \cite{martin2008,aden2009}\\
\end{tabular*}
\caption{Observed parameters and relevant kinematic data for faint MW satellites.}
\end{table*}
}
\newpage

\begin{table*}
\begin{tabular*}{\textwidth}{l|rrrrrrc}
                  & $L_V$ & $c$ & $r_c$ (pc) & $\nu$ & $\alpha$ & $p$ & $\sigma_{int}$ (km\,s$^{-1}$) \\
\hline
\\
Segue             &     340&      0.44&    43.8&    1.74 &   0.58 &   1.22 &   $0.12 \pm 0.01$\\
Segue II          &     860&      0.60&    39.6&    4.13 &   0.69 &   1.42 &   $0.15 \pm 0.02$  \\
Willman I         &    1000&      0.63&    27.7&    4.61 &   0.71 &   1.47 &   $0.19 \pm 0.02$  \\
Bootes II         &    1000&      0.63&    56.5&    4.61 &   0.71 &   1.47 &   $0.13^{+0.01}_{-0.02}$  \\
Coma Berenices    &    3700&      0.86&    62.5&    8.45 &   0.84 &   1.70 &   $0.21 \pm 0.02$  \\
Ursa Major II     &    4100&      0.87&   111.3&    8.75 &   0.84 &   1.71 &   $0.16 \pm 0.02$  \\
Leo V             &    4500&      0.89&    32.7&    9.06 &   0.85 &   1.72 &   $0.31 \pm 0.03$  \\
Canes Venatici II &    7800&      0.99&    51.3&   10.95 &   0.87 &   1.78 &   $0.31 \pm 0.03$  \\
Leo IV            &    8600&      1.00&    78.8&   11.27 &   0.88 &   1.78 &   $0.26 \pm 0.03$  \\
Ursa Major I      &   14000&      1.08&   196.8&   12.95 &   0.90 &   1.82 &   $0.19 \pm 0.02$  \\
Bootes I          &   28000&      1.21&   128.9&   15.84 &   0.92 &   1.86 &   $0.32^{+0.03}_{-0.04}$  \\
Hercules          &   37000&      1.26&   165.9&   16.99 &   0.92 &   1.87 &   $0.32^{+0.03}_{-0.04}$  \\
\end{tabular*}
\caption{Derived characteristics of faint MW satellites assuming $\Upsilon_V=1.45$.}
\end{table*}

\newpage
\begin{table*}
{\tiny
\begin{tabular*}{1.\textwidth}{l | rrrr | rrrr | rrrr | rrrr|}
                  & \multicolumn{4}{|c|}{$log_{10}P$ (normal)} & \multicolumn{4}{|c|}{$log_{10}P$ (normal)} & \multicolumn{4}{|c|}{$log_{10}P$ (uniform)} & \multicolumn{4}{|c|}{$log_{10}P$ (uniform)} \\
                  & \multicolumn{4}{|c|}{$e$ (thermal)} &\multicolumn{4}{|c|}{$e$ (circular)} &\multicolumn{4}{|c|}{$e$ (thermal)} &\multicolumn{4}{|c|}{$e$ (circular)} \\
\hline
\multicolumn{1}{r|}{$P_{min}$ (yrs)} & 1 & 0.1 & 1 & 0.1 & 1 & 0.1 & 1 & 0.1 & 1 & 0.1 & 1 & 0.1 & 1 & 0.1 & 1 & 0.1\\
\multicolumn{1}{r|}{$P_{max}$ (yrs)} & 10& 10 & 100 & 100 & 10& 10 & 100 & 100 & 10& 10 & 100 & 100 & 10& 10 & 100 & 100 \\
\hline
& \multicolumn{16}{|c|}{$f_{bin}(P \le 10$\,yrs$)=5\%$}\\
\hline   
Segue               & 3.6&  5.6&  3.9 & 5.5&   4.1 & 6.5 & 4.2&  6.8 &  3.6 & 5.9 & 4.1&  6.3&   3.8&  8.3&  4.0&  8.3\\
Segue II             & 4.8 & 7.2&  6.4 & 7.0&   6.4&  8.1 & 7.8 & 9.2 &  4.8 & 7.3 & 6.4&  7.7&   6.1& 10.1&  8.4 &10.0\\
Willman I            & 2.4 &4.3 &2.9 &4.4 & 2.9 &5.2 &3.4 &5.8 & 2.5 &5.3 &3.0 &5.2 & 2.9 &6.7 &3.0 &6.4\\
Bootes II & 4.7 &5.8 &5.1 &6.3 & 5.0 &6.3 &5.6 &6.8 & 5.0 &6.4 &5.2 &6.5 & 4.7 &6.9 &5.6 &7.0\\
Coma Berenices & 0.3 &1.3 &0.4 &1.2 & 0.1 &1.5 &0.0 &1.6 & 0.2 &1.9 &0.4 &1.8 & 0.1 &2.4 &0.1 &2.4\\
Ursa Major II & 0.2 &0.8 &0.4 &0.7 & 0.1 &1.0 &0.1 &0.9 & 0.2 &0.9 &0.2 &1.3 & 0.0 &1.1 &0.1 &1.2\\
Leo V & 5.5 &8.0 &6.7 &9.0 & 6.8& 10.0 &8.7 &10.6 & 6.2 &8.7 &6.9 &9.4 & 7.1& 10.7 &7.7 &11.0\\
Canes Venatici II &0.5 &1.8 &0.7 &1.8 & 0.3 &2.2 &0.5 &2.3 & 0.5 &2.4 &0.6 &2.0 & 0.3 &3.2 &0.4 &3.1\\
Leo IV & 21.4& 24.4& 23.5& 24.9 &22.9 &26.0& 24.5& 27.1 &22.3& 25.0& 23.0& 25.3 &23.1& 27.7& 23.9& 27.3\\
Ursa Major I &0.0 &0.0 &0.0 &0.0 & 0.0 &0.0 &0.0 &0.0 & 0.0 &0.0 &0.0 &0.0 & 0.0 &0.0 &0.0 &0.0\\
Bootes I & 0.0 &0.2 &0.0 &0.2 & 0.0 &0.2 &0.0 &0.2 & 0.0 &0.1 &0.1 &0.2 & 0.0 &0.3 &0.0 &0.2\\
Hercules & 1.4 &3.2 &1.8 &3.2 & 1.5 &4.6 &2.0 &4.6 & 1.7 &3.4 &1.8 &3.8 & 2.0 &4.9 &1.9 &5.4\\

\hline
& \multicolumn{16}{|c|}{$f_{bin}(P \le 10$\,yrs$)=15\%$}\\
\hline   
Segue                  & 7.7 & 14.1 &  8.8 & 15.4 &  8.6 & 18.0 & 10.4 & 19.7 &  8.6 & 15.8 &  9.7 & 17.4 &  9.4 & 21.5 & 10.5 & 22.1\\
Segue II               & 13.0 & 18.8 & 17.0 & 21.3 & 17.2 & 23.7 & 21.4 & 25.4 &  13.2 & 19.6 & 16.6 & 21.2 & 17.5 & 25.4 & 21.0 & 27.7\\
Willman I              & 6.8 & 12.0 &  8.4 & 12.8 &  7.6 & 16.6 &  9.4 & 16.5 &  6.7 & 14.3 &  8.4 & 14.6 &  8.6 & 18.9 &  9.3 & 19.1\\
Bootes II              & 8.0 & 11.1 &  8.6 & 11.8 &  8.4 & 13.4 &  9.6 & 14.5 &  7.6 & 12.0 &  8.9 & 12.2 &  8.8 & 14.8 &  9.5 & 15.2\\
Coma Berenices         & 1.2 &  5.8 &  1.7 &  6.0 &  0.5 &  7.9 &  0.6 &  8.4 &  1.5 &  7.8 &  1.7 &  8.3 &  0.6 & 11.6 &  0.6 & 11.9\\
Ursa Major II          & 0.7 &  2.9 &  0.8 &  2.9 &  0.1 &  3.1 &  0.2 &  3.6 &  0.5 &  3.9 &  0.6 &  3.9 &  0.1 &  4.6 &  0.2 &  4.9\\
Leo V                  &13.0 & 19.8 & 16.6 & 22.2 & 17.3 & 24.8 & 21.3 & 27.4 & 13.5 & 21.6 & 17.6 & 23.9 & 16.7 & 28.4 & 21.6 & 30.3\\
Canes Venatici II      & 2.3 &  7.2 &  2.9 &  7.6 &  1.3 &  9.7 &  2.3 & 11.3 &  2.5 &  8.8 &  2.7 &  9.1 &  1.8 & 13.3 &  2.3 & 14.0\\
Leo IV                 &29.4 & 36.7 & 33.7 & 38.7 & 31.9 & 41.1 & 37.1 & 43.0 & 29.6 & 38.8 & 32.3 & 38.7 & 32.0 & 43.4 & 36.3 & 45.1\\
Ursa Major I           & 0.0 &  0.1 &  0.0 &  0.1 &  0.0 &  0.1 &  0.0 &  0.1 &  0.0 &  0.2 &  0.0 &  0.2 &  0.0 &  0.2 &  0.0 &  0.1\\
Bootes I               & 0.1 &  0.9 &  0.2 &  1.2 &  0.0 &  1.3 &  0.0 &  1.5 &  0.2 &  1.5 &  0.2 &  1.6 &  0.0 &  2.3 &  0.0 &  2.6\\
Hercules               & 4.9 & 11.1 &  7.4 & 12.8 &  7.3 & 16.1 &  9.2& 17.9 &  5.4 & 13.5 &  7.5 & 13.9 &  7.5 & 20.0 &  8.5 & 20.8\\                  

\hline
& \multicolumn{16}{|c|}{$f_{bin}(P \le 10$\,yrs$)=30\%$}\\
\hline   
Segue                  &  14.7 & 27.9 & 18.7 & 30.9 & 17.8 & 37.2 & 22.6 & 40.0 & 16.2 & 33.0 & 19.4 & 34.6 & 18.2 & 43.1 & 22.8 & 43.6\\
Segue II               &  24.4 & 34.0 & 33.9 & 39.4 & 33.3 & 43.2 & 41.9 & 46.8 & 25.5 & 37.0 & 31.6 & 40.7 & 31.3 & 45.7 & 40.3 & 48.7\\
Willman I              &  12.9 & 24.8 & 16.5 & 26.1 & 15.5 & 30.8 & 19.7 & 34.0 & 14.0 & 27.7 & 16.9 & 29.5 & 16.5 & 37.5 & 20.7 & 38.2\\
Bootes II              &  13.0 & 18.6 & 14.9 & 19.6 & 14.0 & 23.3 & 16.6 & 24.2 & 12.0 & 20.6 & 13.9 & 21.8 & 14.3 & 25.4 & 16.2 & 27.3\\
Coma Berenices         &   4.2 & 17.6 &  6.1 & 19.2 &  2.5 & 26.3 &  4.7 & 28.3 &  4.7 & 22.6 &  6.0 & 24.4 &  2.8 & 34.9 &  4.5 & 36.8\\
Ursa Major II          &   1.7 &  7.1 &  2.2 &  7.5 &  0.7 &  9.0 &  0.8 & 10.0 &  1.8 &  9.6 &  2.1 &  9.7 &  0.6 & 12.8 &  0.8 & 12.9\\
Leo V                    &  24.5 & 36.0 & 31.9 & 40.8 & 31.6 & 44.1 & 41.3 & 49.7 & 26.1 & 40.0 & 32.0 & 43.7 & 31.5 & 48.6 & 39.7 & 52.1\\
Canes Venatici II      &   6.3 & 18.9 &  9.1 & 21.5 &  6.4 & 27.1 &  8.8 & 30.4 &  6.8 & 23.3 &  8.7 & 24.2 &  6.2 & 35.1 &  9.0 & 36.5\\
Leo IV                 &  40.4 & 52.4 & 47.3 & 54.7 & 44.4 & 58.8 & 52.7 & 63.2 & 41.2 & 55.2 & 45.0 & 57.9 & 45.2 & 63.3 & 51.2 & 65.2\\
Ursa Major I           &   0.0 &  0.5 &  0.0 &  0.5 &  0.0 &  0.4 &  0.0 &  0.7 &  0.0 &  1.1 &  0.0 &  1.1 &  0.0 &  1.1 &  0.0 &  1.2\\
Bootes I               &   0.6 &  3.9 &  0.9 &  4.5 &  0.1 &  6.4 &  0.1 &  7.4 &  0.6 &  6.0 &  0.6 &  6.2 &  0.1 & 10.0 &  0.1 & 10.8\\
Hercules               &  12.8 & 25.5 & 19.8 & 31.1 & 18.6 & 38.7 & 25.4 & 42.1 & 14.1 & 31.7 & 19.1 & 33.1 & 19.2 & 44.1 & 24.8 & 46.9\\
\end{tabular*}
\caption{Simulation results for each satellite
  assuming $f_{bin}(P\le10$\,yrs$)=5\%,~15\%$ and
  $30\%$ (upper, middle and lower sections, respectively), for
  different period and eccentricity assumptions. Numbers
  refer to the percentage of simulations that generate a velocity
  dispersion that is greater than or equal to the 1$\sigma$ lower
  error-bound on the observed velocity dispersion.}
}
\end{table*} 

\newpage

\begin{figure*}
  \begin{center}
    \includegraphics[angle=270, width=5.95cm]{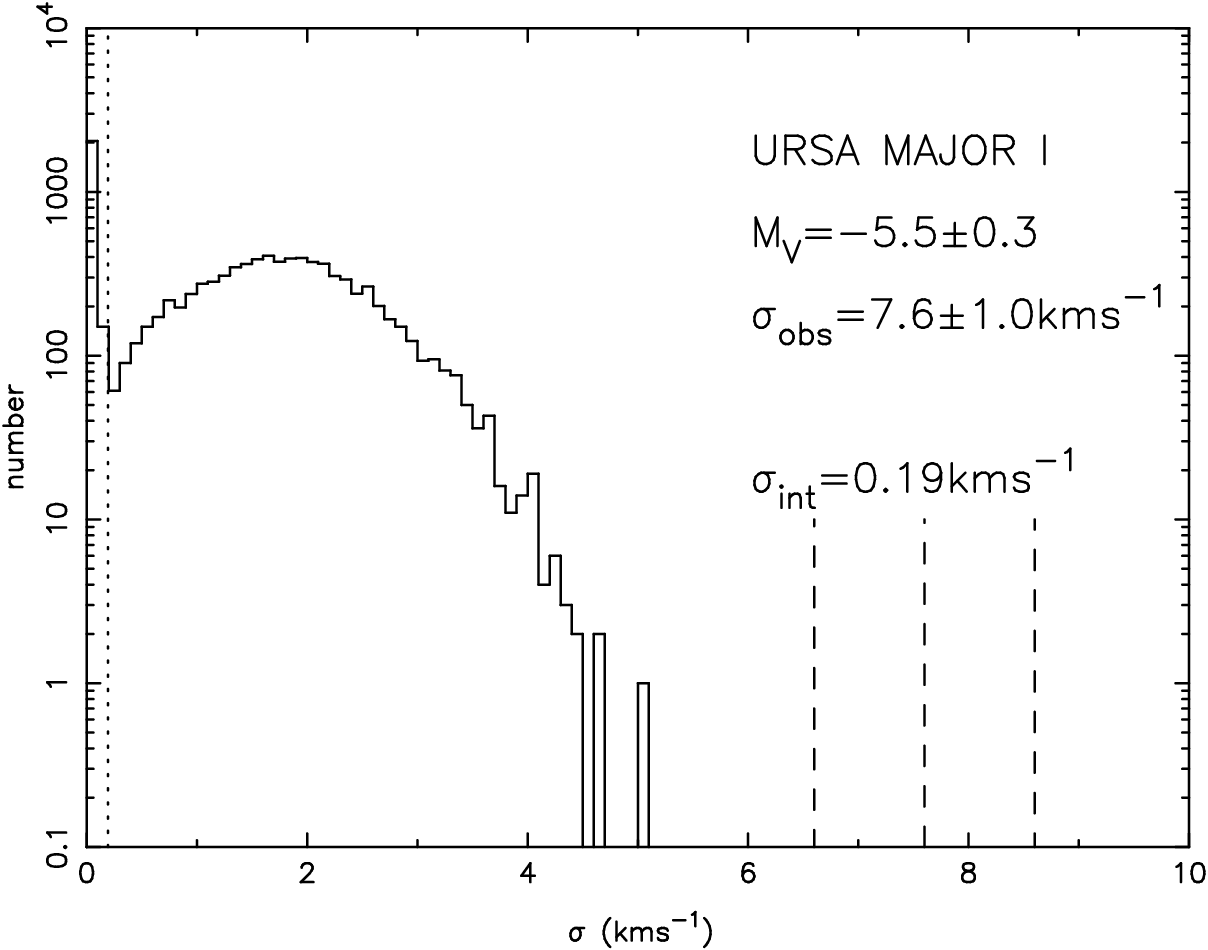}
    \includegraphics[angle=270, width=5.95cm]{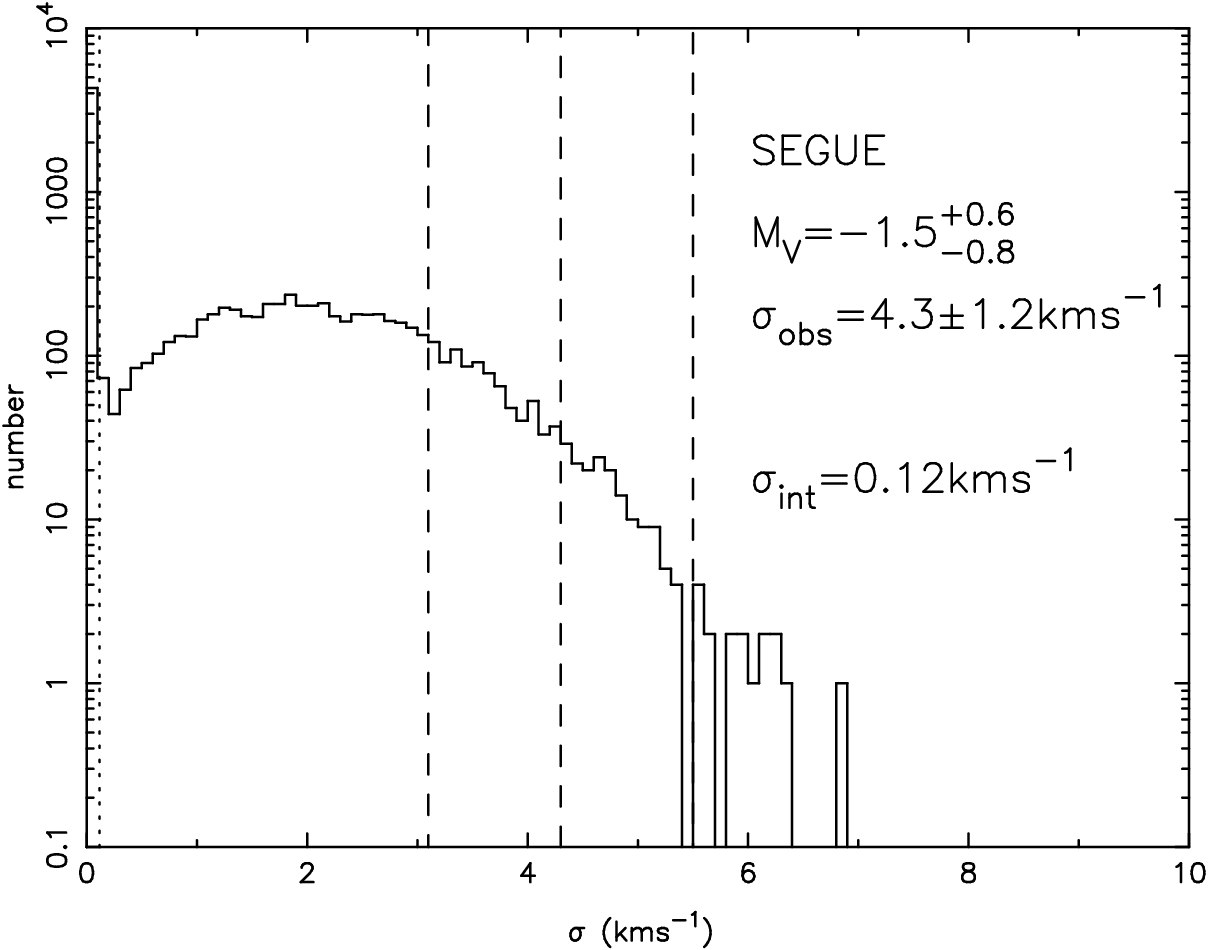}
    \includegraphics[angle=270, width=5.95cm]{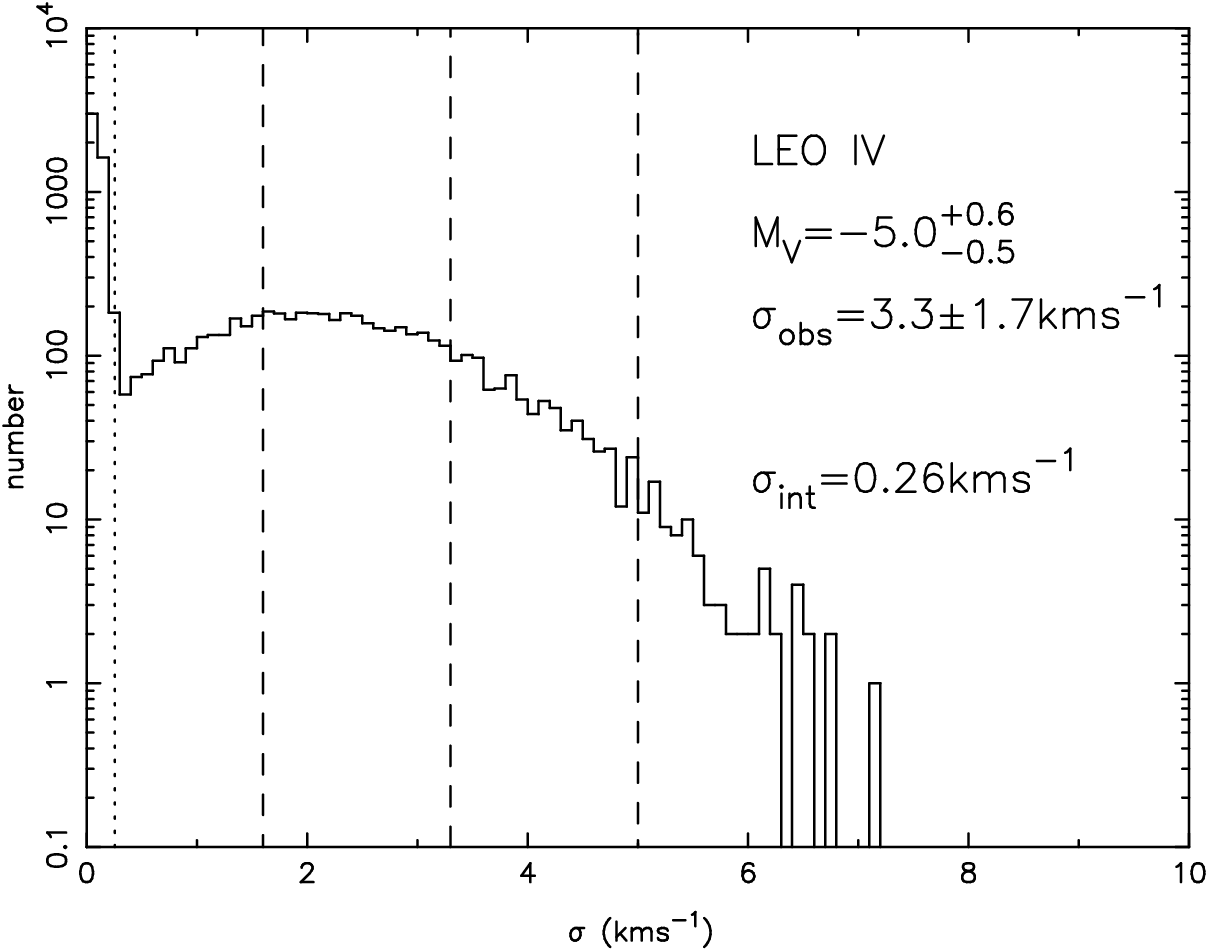}
    \caption{The simulated velocity dispersion distribution for Ursa
      Major~I, Segue, and Leo~IV, assuming $f_{bin}(P\le
      10$\,yrs$)=15\%$, a log-normal period distribution with $1\le
      P\le 100$\,yrs and circular orbits (column~7 of Table~3). The
      dotted line shows the assumed intrinsic velocity dispersion, and
      the dashed lines show the observed velocity dispersion with
      1$\sigma$ error bounds. Ursa Major~I is incompatible with an
      intrinsically low mass system, but this hypothesis cannot be
      excluded at high confidence for Segue and Leo~IV.}
  \end{center}
\end{figure*}

\newpage

\begin{figure*}
  \begin{center}
    \includegraphics[angle=270, width=15cm]{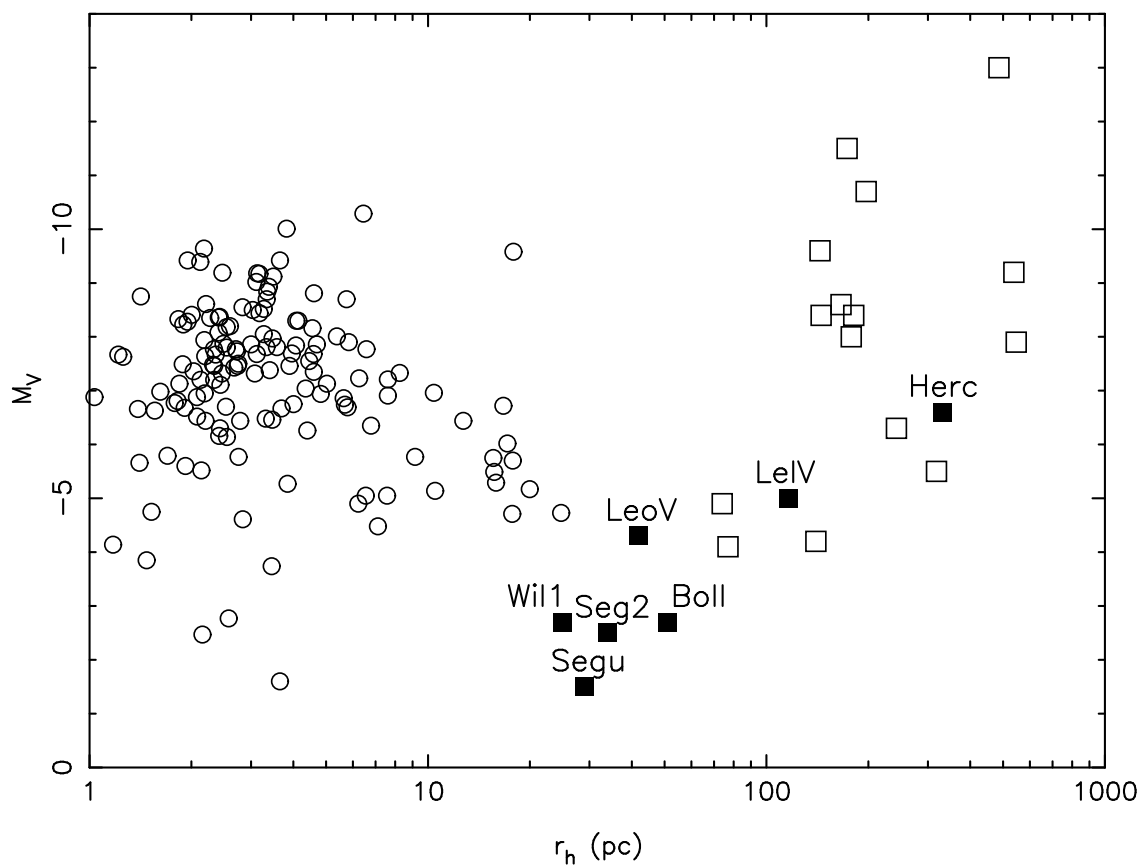}
    \caption{$r_h$ versus $M_V$ for MW satellites. Open circles
      correspond to globular clusters (\citealt{harris1996}), squares
      correspond to dwarf galaxies and dwarf candidates. The
      putative dwarfs whose velocity dispersions are most
      susceptible to boosting by spectroscopic binaries are
      shown as filled squares. These include all dwarf
      candidates with $r_h\lesssim70$\,pc that lie closest to
      the globular cluster population.}
  \end{center}
\end{figure*}

\end{document}